\documentclass[apl,twocolumn,showpacs,amsmath,amssymb,floatfix]{revtex4}
\usepackage{graphicx}
\usepackage{dcolumn} 
\usepackage{bm}      
\usepackage{subfigure}

\newcommand{\be}{\begin{equation}}
\newcommand{\ee}{\end{equation}}
\newcommand{\bea}{\begin{eqnarray}}
\newcommand{\eea}{\end{eqnarray}}
\newcommand{\bw}{\begin{widetext}}
\newcommand{\ew}{\end{widetext}}
\newcommand{\mm}{\mathrm}

\newcommand{\ud}{\mm{d}}

\newcommand{\bi}{\begin{itemize}}
\newcommand{\ei}{\end{itemize}}

\newcommand{\half}{\frac{1}{2}}

\begin{document}

  \title{The role of dipolar interactions for the determination of intrinsic switching field distributions}

  \author{Yang Liu}
  \author{Karin A.\ Dahmen}
  \affiliation{Department of Physics, University of Illinois at
    Urbana-Champaign, Urbana, IL 61801, USA}
  \author{A. Berger}
  \affiliation{CIC nanoGUNE Consolider, E-20009 Donostia - San Sebastian, Spain}

  \date{\today}

  \begin{abstract}

The $\Delta H(M, \Delta M)$ method and its ability to determine 
intrinsic switching field distributions of perpendicular recording media are
numerically studied. It is found that the presence of 
dipolar interactions in the range of typical recording media substantially 
enhances the reliability of the $\Delta H(M,\Delta M)$ method. In 
addition, a strong correlation is observed between the precision of this 
method and a self-consistency-check of the data sets, which is based upon a 
simple redundancy measure. This suggests that the latter can be utilized as 
an efficient criterion to decide if a complete data analysis is warranted or 
not. 

  \end{abstract}


  \maketitle


For magnetic recording media used in state-of-the-art storage applications 
such as hard disks drives, the intrinsic switching-field distribution $D(H_\mm{S})$ 
of the media grains is one of the most crucial properties defining the 
recording quality~\cite{Shimizu-03}. In general, each grain is characterized by an 
intrinsic switching field $H_\mm{S}$, which is a local material property. Because 
grains interact with each other by means of exchange and dipolar 
interactions, $D(H_\mm{S})$ is not easily accessible in macroscopic measurements. 
This is especially true for perpendicular recording media due to the 
strength of the interactions, which is much larger than in the previously 
used longitudinal recording media~\cite{Berger-05}. Therefore, many attempts to determine 
$D(H_\mm{S})$ from macroscopic magnetization reversal type measurements have had 
varying success with none of the presently available methodologies being 
completely satisfactory for all levels of intergranular
interactions~\cite{Shimizu-03,Berger-05, Tagawa-91, Pike-99, Veerdonk-02,
  Berger-06,Winklhofer-06,Berger-07}. 

The recently developed $\Delta H (M, \Delta M)$ method has been 
used in analyzing and quantifying progress in perpendicular 
recording media fabrication~\cite{Berger-05,Berger-06,Berger-07}. This method measures the field 
difference $\Delta H$ at constant magnetization $M$ between the major 
hysteresis loop and a number of recoil curves, which each start at a certain 
distance $\Delta M$ away from saturation. Within the mean-field 
approximation, the functional dependency of $\Delta H$ can be written as 
$\Delta H(M,\Delta M) = I^{-1}((1-M)/2) - I^{-1}((1-M-\Delta M)/2)$
where $I^{-1}$ is the inverse of the integral $I(x) = \int_{-\infty}^{x} D(H_\mm{S}) \, \ud
  H_\mm{S}$. Within the framework of this method, $\Delta H$ is 
independent from the grain interactions, which allows for a direct 
experimental access to determining $D(H_\mm{S})$. For certain parameterized 
distribution functions, one can derive analytic expressions for $\Delta H$. 
For example, for a Gaussian distribution of width $\sigma$, one finds
$\Delta H_\mm{G}(M,\Delta M) =  \sqrt{2} \sigma [\mm{erf}^{-1}(M+\Delta
  M) - \mm{erf}^{-1}(M) ]$. Details of this method and the analysis formalism have 
been described previously~\cite{Berger-05,Berger-06,Berger-07,Liu-07}.

The $\Delta H(M, \Delta M)$ method was demonstrated to have 
several advantages over comparable methods. First, it allows for the 
determination of the entire $D(H_\mm{S})$ distribution and its functional form and 
not just a single characteristic parameter~\cite{Berger-05,Berger-06}. Second, it has a 
well-defined reliability range and it allows for oversampling, which makes 
self-consistency checks feasible~\cite{Liu-07}. Third, its failure mode was found to 
show universal behavior, independent from the detailed lattice structure in 
numerical simulations~\cite{Liu-08}.

Despite these advantages, it is still unknown whether the reliability range 
of the $\Delta H (M, \Delta M)$ method with respect to exchange 
interactions might be affected by the simultaneous presence of dipolar 
interactions. Given the fact that this simultaneous presence of both 
interactions is the realistic case for actual recording media, it is an 
important issue, which we have studied in this letter.

For our numerical studies, we model each media grain as a symmetric 
hysteron, which generates a rectangular hysteresis loop in an applied field 
$H$~\cite{Liu-07}. The half width of the hysteresis loop is just the intrinsic switching 
field $H_\mm{S}$ of this hysteron. We further assume that the magnetization of each 
hysteron exhibits values of $\pm 1$ only and orients exactly along the applied 
field $H$. The ferromagnetic layer system is then represented by a square or 
triangular lattice of symmetric hysterons with periodic boundary conditions. 
The model Hamiltonian can then be written as ${\cal H} = - J_\mm{ex} \sum_{{<}i,j{>}} S_i S_j + J_\mm{dp}
  \sum_{i\neq j} \frac{S_i S_j}{r^3_{ij}} - \sum_i ( H + \mm{sgn}(S_i)
  {H_\mm{S}}_i ) S_i$. Here, the first term represents that hysterons interact 
ferromagnetically with their nearest neighbors by means of exchange 
interactions of strength $J_\mm{ex}$. The second term represents that hysterons 
exhibit a distance-dependent dipolar interaction of strength $J_\mm{dp}$ with 
all other hysterons. The third term accounts for the effect of the external 
field and the intrinsic switching field. This model is referred to as the 
\textit{interacting random hysteron model}. The algorithm for the simulation of the major hysteresis loops and the 
corresponding recoil curves of this model is described in~\cite{Liu-07}. For the 
numerical calculations of long-range dipolar interactions in a system with 
periodic boundary conditions, we utilized the efficient formalism described 
by Lekner~\cite{Lekner-91}.

For our numerical study of the $\Delta H(M, \Delta M)$ method's 
reliability, we assume a Gaussian distribution $D(H_\mm{S})$ of width $\sigma=1$ 
for a two-dimensional square lattice comprising of total $N$ hysterons. 
Different system sizes ranging from $50^2$ to $400^2$ have been studied 
to estimate finite-size inaccuracies. Results presented here were calculated 
for $100^{2}$, which we found to be sufficiently precise in all cases. In 
our simulations, we normalize all parameters to $\sigma$ and vary both 
$J_\mm{ex}$ and $J_\mm{dp}$. For each parameter set ($J_\mm{ex}$, $J_\mm{dp}$), we 
calculate the complete set of $M(H)$-curves (both the saturation hysteresis 
loop and recoil curves), from which $\Delta H(M,\Delta M)$ data sets are 
then extracted.

  \begin{figure}[t]
    \includegraphics[width=0.5\textwidth]{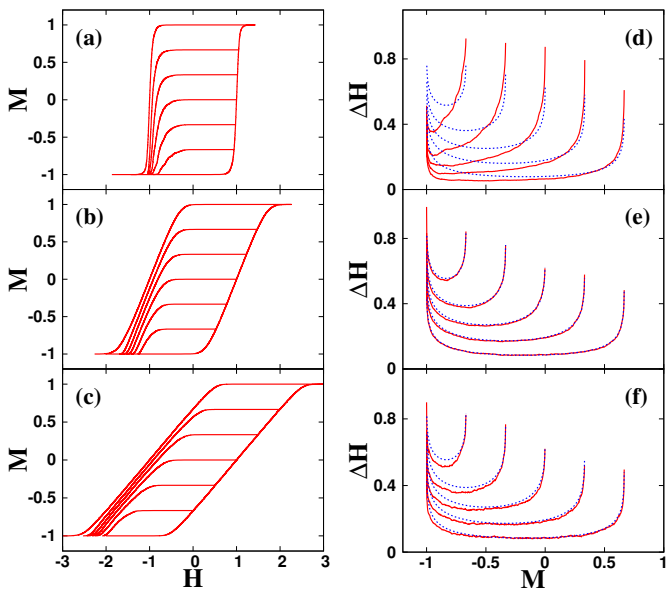}
    \caption{\label{fig:MH} (Color online) Using Gaussian $D(H_\mm{S})$ with
      width $\sigma=1.0$ to calculate the $M(H)$ curves and $\Delta H(M,\Delta
      M)$ curves on a 2D square lattice with $N=100^2$ hysterons and
      $J_\mm{ex}=0.4$. (Top) $J_\mm{dp}=0$. (Middle) $J_\mm{dp}=0.4$. (Bottom)
      $J_\mm{dp}=0.8$. (Left) $M(H)$ curves: main loop and 5 recoil
      curves. (Right) $\Delta H(M, \Delta M)$ curves for the 5 
      recoil curves: (solid lines) numerical result; (dotted lines) mean-field
      approximation. Here $M$ (or $\Delta M$) is normalized to the
      saturation value $M_\mm{S}=N$ and $H$ (or $\Delta H$) is normalized to the
      coercive field $H_\mm{C}$.}
  \end{figure}

The results displayed in Fig.1 show several specific examples for parameter 
sets $(J_\mm{ex}, J_\mm{dp})$ = (0.4,0), (0.4,0.4) and (0.4,0.8) . The simulated 
$M(H)$ curves are shown in the left column. It is clearly seen that increasing 
the strength of dipolar interactions shears the hysteresis loops 
substantially as expected. The right column displays the corresponding 
$\Delta H(M,\Delta M)$ curves. The solid lines are the numerically 
extracted results from the simulated $M(H)$ curves while the dotted lines 
denote the mean-field behavior according to the expression of $\Delta H_\mm{G}
(M,\Delta M)$. Comparing Fig.1(d), (e) and (f), we find that  
dipolar interactions of intermediate strength make the system most 
mean-field like.

To study the effect of dipolar interactions on the $\Delta H(M,\Delta M)$
method's reliability in a more systematic and quantitative way, we need to
introduce quantitative reliability measures. Obviously, the  
reliability range of the mean-field approximation, upon which the 
$\Delta H (M, \Delta M)$ method is based, can be checked by means 
of a least-squares fit of $\Delta H_\mm{G} (M, \Delta M)$ to the numerical 
data and a subsequent analysis of the conventional fit-quality measures, 
such as: (1) the square of the multiple correlation coefficient $R^{2}$ and 
(2) the percentage difference $P_\mm{d}$ between the fitting result and the 
input parameter $\sigma$. By definition, $R^2=1$ and $P_\mm{d}=0$ would 
correspond to perfect data fitting, i.e. the exactness of the mean-field 
limit. We therefore calculated a least-squares fit to $\Delta
H_\mm{G}(M,\Delta M)$ for the numerically extracted $\Delta H(M,\Delta M)$
data for each parameter set of ($J_\mm{ex}, J_\mm{dp}$). From  
these fits, we then computed both $P_\mm{d}$ and $R^2$. The results are shown 
in contour plots (see Fig.2).

The shape of the contour plots is rather interesting. First, it is nearly 
symmetric along the diagonal direction, i.e. $J_\mm{dp}/J_\mm{ex}=1$. This 
clearly demonstrates that the roles of exchange and dipolar interactions in 
determining $D(H_\mm{S})$ are almost equally important. Individually increasing 
either one will make the $\Delta H(M, \Delta M)$ method less 
reliable, while increasing both of them with proper strength ratio of order 
1 will substantially extend the reliability range. Second, the shape is not 
really symmetric. It is tilted upwards and smoother on the high $J_\mm{dp}$ 
side than the high $J_\mm{ex}$ side. This suggests that the $\Delta H(M, \Delta
M)$ method can clearly cope with higher dipolar interactions  
than exchange interactions, in agreement with previous micromagnetic tests 
~\cite{Berger-06}.

 \begin{figure}[t]
   \includegraphics[width=0.50\textwidth]{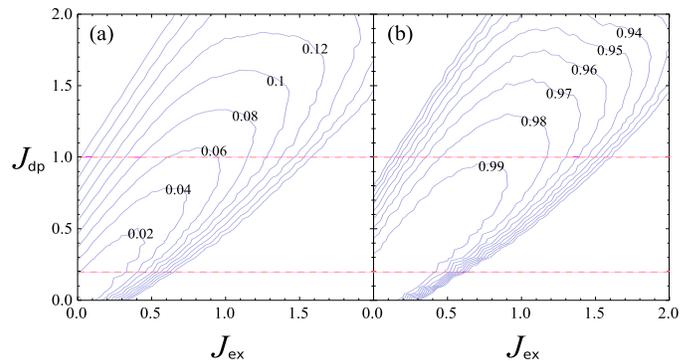}
    \caption{\label{fig:reliability-Je-Jd-contour}
 (Color online) Contour plots of the fit quality measures as functions of
 $J_\mm{ex}$ and $J_\mm{dp}$. (a) $P_\mm{d}$. (b) $R^2$. Dotted lines indicate
 the range of $J_\mm{dp}$ for realistic materials.}
\end{figure}

  \begin{figure}[t]
    \includegraphics[width=0.45\textwidth]{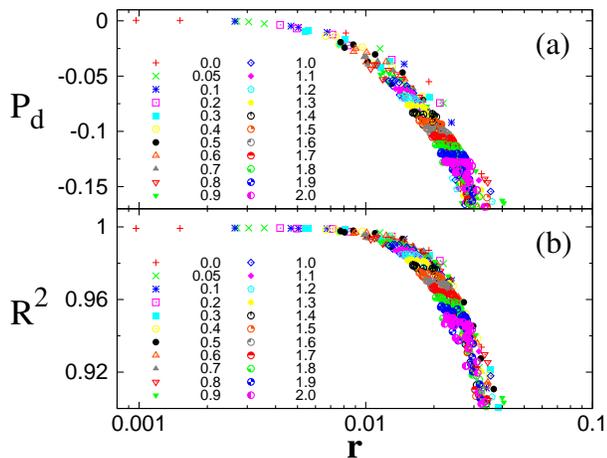}
    \caption{\label{fig:correlation} (Color online) Correlations between the reliability
      measures: (a) $P_\mm{d}$ and $r$; (b) $R^2$ and $r$. Each data point
      represents a different parameter set ($J_\mm{ex}$, $J_\mm{dp}$). Data
      points with the same $J_\mm{dp}$ values are grouped and shown with the
      same point symbol, as indicated by the legend.}
 \end{figure}

The overall shape of the contours can be qualitatively explained by the 
\textit{interaction compensation effect}. As we know, the intergranular exchange interactions are ferromagnetic (FM) 
and short-range while the dipolar interactions are anti-ferromagnetic (AFM) 
and long-range. The competition between the two ``opposite'' interaction 
tendencies will yield a variety of system behaviors. Generally speaking, as 
we steadily increase $J_\mm{ex}$ from 0 to higher values while keeping $J_\mm{dp}$ 
constant, we shift the system from the AFM-interaction-dominated regime to 
the mean-field regime and to the FM-interaction-dominated regime. Only 
within the interaction compensation region, the FM and AFM interaction 
tendencies nearly cancel each other. Consequently, the system is most 
mean-field like and the $\Delta H (M, \Delta M)$ method becomes 
most reliable there. From the model Hamiltonian, we notice that if 
$J_\mm{ex}/J_\mm{dp}=1$, the exchange and dipolar interactions will cancel 
exactly for the nearest-neighboring hysterons in the lattice. However, due 
to the long-range and distance-dependent features of dipolar interactions, 
this cancellation will not be exact for hysterons with longer distances. 
This explains why the interaction compensation region is only roughly 
symmetric along the diagonal direction. 

To quantify the interaction compensation region or equivalently the 
reliability range of the $\Delta H (M, \Delta M)$ method, one can 
define a critical value for each reliability measure, above which this 
method is sufficiently accurate. For example, we might define 
$R^{2}_\mm{c}=0.98$ to be the critical value for $R^2$ in accordance with 
the best available experimental data~\cite{Berger-06}. Then the reliability range for the 
parameter set ($J_\mm{ex}, J_\mm{dp}$) can be clearly seen from the region 
enclosed by the second highest contour in the $R^{2}$ plot. This particular 
contour will be referred to as the critical contour, inside which the system 
is virtually mean-field like and the $\Delta H(M, \Delta M)$ 
method is reliable. Note that for practical recording media, the ratio of 
$J_\mm{dp}/\sigma$ will probably be limited within the range of $0.2 \sim 1$ as 
illustrated in Fig.2. One can easily see that within the realistic 
$J_\mm{dp}/\sigma$ range, the dipolar interactions improve the reliability 
range of the $\Delta H(M, \Delta M)$ method up to higher $J_\mm{ex}$ 
values. We also notice that for $J_\mm{ex}/\sigma \le 0.2$ (read off from the 
intercept of the $R^{2}$ critical contour on the $J_\mm{ex}$-axis), higher 
dipolar interaction will only make the $\Delta H(M, \Delta M)$ method
worse. But this part is very small compared to the part where dipolar
interactions make the $\Delta H(M, \Delta M)$ method robust.

Besides the fit-quality measures $R^2$ and $P_\mm{d}$, there is a 
self-consistency-check measure, which is based upon data redundancy in 
between multiple recoil curves~\cite{Liu-07}. One can test data for deviations from 
this redundancy by means of a quantity $r=\frac{1}{n}\sum_{i,j} \left\langle
  r^2_{ij}(M) \right\rangle^{\half}$ where $r_{ij}(M)$ is by definition identical to zero within the mean-field 
approximation, so is $r$~\cite{note}. The specific advantage of this quantity $r$ is 
that it can be directly calculated from data sets alone without the need for 
any data fitting. Therefore, it is important to analyze the possible 
correlation between the fit-quality measure (either $R^2$ or $P_\mm{d}$) and 
the deviation-from-redundancy measure $r$. Knowledge of this correlation will 
enable us to estimate the suitability of the $\Delta H(M,\Delta M)$ method 
without any data fitting. Considering this, we calculated $r$ from the 
numerical $\Delta H(M,\Delta M)$ data for each parameter set of ($J_\mm{ex}$, 
$J_\mm{dp}$). Overall, the contour plot of $r$ shows very similar features as of 
the ones being displayed in Fig.2 for $R^2$ and $P_\mm{d}$. To visualize and 
quantify the correlation between $R^2$ ($P_\mm{d}$) and $r$, we plot $R^2$ 
($P_\mm{d}$) vs. $r$ for the complete set of different ($J_\mm{ex}$, $J_\mm{dp}$) 
parameter (see Fig.3). We find that the data collapse fairly well onto a 
single line in the high $R^2$ or low $|P_\mm{d}|$ range, in 
which the utilization of the $\Delta H (M, \Delta M)$ method is 
sensible and accurate. This indicates that $R^2$ and $P_\mm{d}$ are highly 
correlated with $r$ in the regime where these quantities matter. Due to the 
knowledge of these correlations, one now has a criterion that enables a 
judgment on the usefulness and reliability of any $\Delta H(M,\Delta 
M)$-data set evaluation. For that, one simply determines the $r$ value from 
experimental or modeling data sets, looks up the expected precision with the 
help of Fig.3 and then decides if a further data analysis is warranted or 
not.

In summary, we find that the presence of dipolar interactions similar in 
size to those of real perpendicular recording media makes the 
$\Delta H (M, \Delta M)$ method substantially more precise and 
robust. The deviation-from-redundancy measure $r$, which is a 
self-consistency-check, is found to be a good predictor of the 
$\Delta H (M, \Delta M)$ method's reliability and can be utilized 
as a criterion to decide if a full scale data-analysis is warranted.

Y.L. and K.D. acknowledge support from NSF 
through grants DMR 03-14279 and the Materials Computation Center DMR 
03-25939(ITR). Work at nanoGUNE acknowledges funding from the Department of 
Industry, Trade and Tourism of the Basque Government and the Provincial 
Council of Gipuzkoa under the ETORTEK Program, Project IE06-172, as well as 
from the Spanish Ministry of Science and Education under the 
Consolider-Ingenio 2010 Program, Project CSD2006-53.


\begin{thebibliography}{99}
\bibitem{note} For recoil curve pair $(i,j)$,  the
  deviation from redundancy is defined as $r_{ij} (M)=\frac{\Delta H_i
    (M)+\Delta H_j (M-\Delta M_j )-\Delta H_i  
(M-\Delta M_j )-\Delta H_j (M-\Delta M_j +\Delta M_i )}{\Delta H_i 
(M)+\Delta H_j (M-\Delta M_j )+\Delta H_i (M-\Delta M_j )+\Delta H_j 
(M-\Delta M_j +\Delta M_i )}.$ In mean field theory, it can be exactly proven 
that $r_{ij}(M)=0$. For details, see Ref~\cite{Liu-07}.
\end{thebibliography}
\end{document}